\documentstyle[12pt,proc,epsfig]{article}

\begin{document}
\onecolumn
\begin{flushright}
TUHEP-TH-00118\\
hep-ph/0201167
\end{flushright}
\begin{center}{\bf PROBING THE ELECTROWEAK SYMMETRY BREAKING MECHANISM \\AT
HIGH ENERGY COLLIDERS}
\end{center}
\baselineskip=16pt
\vspace{0.5cm}
\centerline{\rm YU-PING KUANG}
\baselineskip=13pt
\centerline{\it Department of Physics, Tsinghua University,}
\baselineskip=12pt
\centerline{\it Beijing, 100084, China}
\vspace{0.2cm}
\baselineskip=13pt
\vspace{0.6cm}
\abstract{We briefly review the recent developments of probing the
electroweak symmetry breaking mechanism at high energy colliders such
as the CERN LEP2, the Fermilab Tevatron, the CERN LHC and the $e^+e^-$ linear 
colliders. Both weakly interacting and strongly interacting electroweak 
symmetry mechanisms are concerned.}
\null\vspace{0.2cm}
\baselineskip=14pt
\section{Introduction}\vspace*{-0.4cm}

It is remarkable that the electroweak standard model (EWSM) has successfully 
passed all the precision tests. However, despite of the present success, the 
electroweak symmetry breaking mechanism (EWSBM) is not clear yet.
All results of the experimental searches for the Higgs boson 
are negative. So far, we only know the existence of a vacuum expectation value 
(VEV) $v=240$ GeV which breaks the electroweak gauge symmetry, but we donnot 
know if it is just the VEV of the elementary Higgs boson in the EWSM or not, 
and we even donnot know if there is really a Higgs boson below 1 TeV. The 
unclear EWSBM is a big puzzle in particle physics, and the probe of the EWSBM 
is one of the most important problems in current high energy physics.
Since all particle masses come from the VEV $v$, probing the EWSBM 
concerns the understanding of the {\it origin of all particle 
masses}, which is a very fundamental problem in physics. 
The latest experimental bound of the Higgs boson mass given by the LEP Working 
Group for Higgs Boson Searches is already $~m_H>107.7$ GeV \cite{LEP-m_H}.
New TeV energy colliders are definitely needed to further study this important 
problem experimentally. 

From the theoretical point of view, there are several unsatisfactory features
in the Higgs sector in the EWSM, e.g. there are so many free parameters related
to the Higgs sector, and there are the well-known problems of {\it triviality}
and {\it unnaturalness} \cite{Chanowitz}. Usually, people take the 
point of view that the present theory of the EWSM is only valid up to a 
certain energy scale $\Lambda$, and new physics beyond the EWSM will become 
important above $\Lambda$. Possible new physics are supersymmetry
(SUSY) and dynamical EWSBM concerning new strong interactions, etc. So that 
probing the mechanism of EWSB also concerns the discovery of new physics. 

In the following, we shall give a brief review of the present developments
of probing the EWSBM at various high energy colliders.

\section{The Higgs Boson}
\subsection{Where is the Higgs Boson?}

In the EWSM, the Higgs boson mass $m_H$ is a free parameter related to the
Higgs self-coupling constant $\lambda$. We now look at some possible hints
of $m_H$ from theoretical and experimental studies.

Let us first look at the theoretical hint. We know that if the EWSM is valid 
in the whole energy range, the renormalized coupling constant 
$\lambda\to 0\--${\it triviality} \cite{Chanowitz}.
Since the Higgs boson develops a nonvanishing VEV only if it has a nontrivial 
self-interaction $\lambda\not=0$, triviality is a serious problem 
of the EWSM. 
To avoid triviality, people usually take the point of view that the EWSM may 
not be fundamental but is a low energy effective theory of a more 
fundamental theory below a certain physical scale $\Lambda\not\to \infty$ 
The scale $\Lambda$ serves as a natural momentum cut-off which is the
highest energy scale in the effective theory.
The problem of triviality can then be avoided if the fundamental theory does 
not suffer from a triviality problem. The larger the scale $\Lambda$
the smaller the nonvanishing coupling $\lambda$.
Note that $m_H$ is proportional to $\lambda$, so that there is an upper bound 
on $m_H$ for a given $\Lambda$ \cite{HR}. A careful calculation of such a 
triviality bound on $m_H$ has been given in Ref. 3 and is shown as the 
upper curve in Fig. 1 \cite{HR}. By definition, $m_H$ cannot exceed the
highest scale $\Lambda$ in the effective theory. This determines the maximal 
value of $m_H$ which is {\it of the order of 1 TeV} \cite{Chanowitz,HR}.

\null\vspace{0.4cm}

\centerline{\epsfig{figure=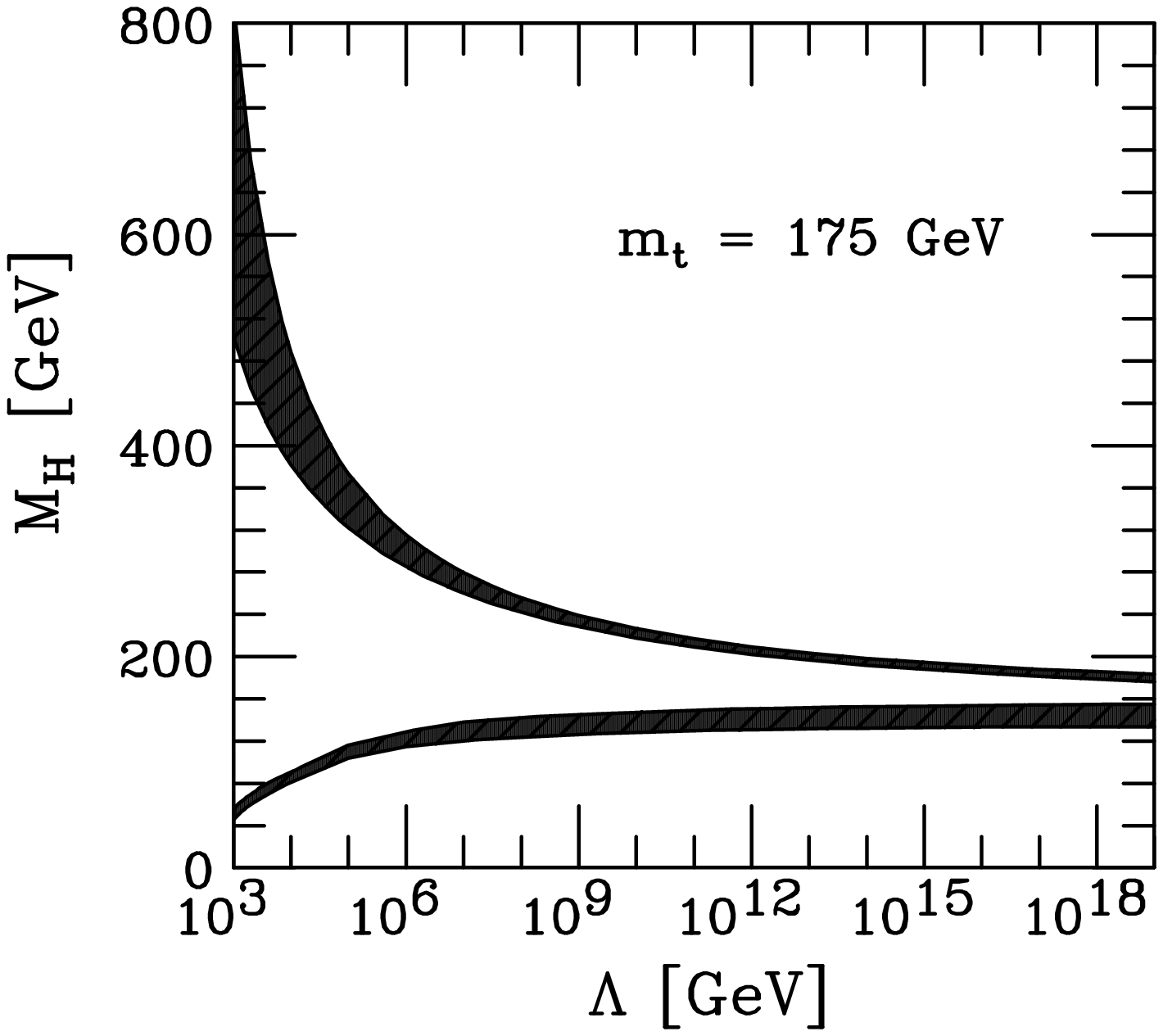,height=5cm,angle=90}}
\null\noindent{\footnotesize {\bf Fig. 1.} The triviality bound (upper curve) and the vacuum
stability bound (lower curve) on $m_H$ in the EWSM.
The solid areas as well as the crosshatched area indicate theoretical 
uncertainties. Quoted from Ref. 3.}
\null\vspace{0.4cm}

On the other hand, when loop contributions are concerned, the stable physical 
vacuum state should be determined by the minimum of the {\it effective 
potential} $V_{eff}$. In $V_{eff}$, the 
Higgs boson loop (with the Higgs self-interaction) tends to 
stabilize the physical vacuum with a nonvanishing $v$, while the fermion loop 
tends to destabilize the physical vacuum \cite{Chanowitz}. The heavier the 
fermion the stronger the violation of vacuum stability. 
The $t$ quark gives a strong violation of the vacuum stability. To obtain 
a stable physical vacuum, a large enough Higgs self-interaction is needed to 
overcome the destabilization from the $t$ quark loop contributions. 
This requirement gives a {\it lower bound on the Higgs mass $m_H$}. The vacuum 
stability bound on $m_H$ is shown as the lower curve in Fig. 1 \cite{HR}.

The region between the two curves in Fig. 1 is the allowed region.
We see that there is a possibility of extrapolating the EWSM up to the Planck 
mass, if and only if the Higgs mass $m_H$ is around 160 GeV. Of course, Fig. 1 
tells nothing about where the actual scale of new physics really is. Even if a 
Higgs boson of $m_H\approx 160$ GeV is found, Fig. 1 still allows $\Lambda$ to 
take any value below the Planck mass. Of special interest is that if a very 
light Higgs boson with $m_H\sim 100$ GeV or a heavy Higgs boson with $m_H\leq 
500$ GeV is found. Then Fig. 1 shows that $\Lambda$ will be at most of the 
order of TeV, and this energy can be reached at the LHC and LCs.
Furthermore, {\it If a Higgs boson is not found below 1 TeV, we should find 
new physics in this region}.

There is an important conclusion for the Higgs boson mass in the minimal SUSY 
extension of the standard model (MSSM). Careful theoretical studies on 
the MSSM Higgs mass up to two-loop calculations show that the mass of the 
lightest CP even Higgs boson $h$ in the MSSM cannot exceed a bound 
$m_h|_{max}\approx 130$ GeV \cite{HHW} with a theoretical uncertainty
about 5 GeV, which can be reached by all the designed LCs. 
If $h$ is not found below 135 GeV, MSSM will be in a bad shape and SUSY models 
beyond the MSSM should be seriously considered. 

Next we look at some possible experimental hints. There are various
analyses of the best fit of the electroweak theory to the LEP/SLD 
data at the $Z$-pole which give certain requirements on the Higgs mass.

\null\noindent{\bf i, Best Fit of SM to the $Z$-pole Experiments}

The high precision of the LEP/SLD data can give certain expected value of 
$m_H$ from the requirement of the best fit. For instance, the analysis
in Ref. 5 shows that the best fit value of $m_H$ is \cite{EL}
\begin{eqnarray}                       
m_H=107^{+67}_{-45}~{\rm GeV}\,.
\label{SMfit}
\end{eqnarray}
The upper bounds of $m_H$ at the $90\%$ C.L. is $m_H<$ 220 GeV \cite{EL}.
These numbers imply that the Higgs boson may be found in the near future
if it exists. It should be noticed that this is the conclusion from an 
analysis using {\it only the pure EWSM formulae without including any effects 
of new physics}.

\null\noindent{\bf ii, Combining the $Z$-pole data and the Direct Search Bound}

Apart from the above hint from the $Z$-pole data, there have been direct 
searches for the Higgs boson at LEP in recent years with negative results.
If one combine the two sources of data, the probability distribution of the 
Higgs mass will change, and the resulting expected value
of the Higgs mass will be different from eq.(\ref{SMfit}). This kind of
study has been carried out in Ref. 6 taking
account of the direct search bound $m_H>89.8$ GeV from the $\sqrt{s}=183$ GeV
run of LEP in 1998. The result is \cite{DD1}
\begin{eqnarray}                              
m_H=170\pm 80~{\rm GeV}\,,~~~~~~~~~~
m_H<300~{\rm GeV}\,,~~95\% ~C.L.
\label{combine}
\end{eqnarray}
An upgraded analysis by the same authors has also been given with similar 
conclusions \cite{DD1}. We see that this expected $m_H$ is higher than that 
obtained merely from the $Z$-pole data. Now the direct search bound has 
increased to $m_H>107.7$ GeV \cite{LEP-m_H} which will make the expected $m_H$ 
further higher.

\null\noindent{\bf iii, Considering New Physics Contribution to $S$}

The above results are all based on analyses using only the pure EWSM 
formulae. Since the EWSM may only be valid below a certain physical scale 
$\Lambda$, new physics may affect the $Z$-pole observables, or the parameters 
$S,~T,~U$ and $\epsilon_b$. Ref. 7 has given an interesting 
with $S$ treated as a new parameter (including unknowm new physics effect), and 
the best fit values of $S,~m_H,~m_t$ and $\alpha_s$ are \cite{Erler}
\begin{eqnarray}                    
S=-0.20^{+0.24}_{-0.17}\,,~~~~m_H=300^{+690}_{-310}\,,
m_t=172.9\pm 4.8~{\rm GeV}\,,~~~~\alpha_s=0.1221\pm 0.0035\,.
\label{Erler}
\end{eqnarray}
The best fit values of $m_t$ and $\alpha_s$ are all close
to the world averaged values, and the expected value of $m_H$ is much 
uncertain (the upper value is of the order of 1 TeV) when the formula for $S$ 
is relaxed. 

\null\noindent{\bf iv, Best Fit of the Electroweak Chiral Lagrangian to the 
$Z$-pole Data}

Another interesting analysis was recently given in Ref. 8. Since the Higgs 
boson is not found, the authors consider the possibility that there is no 
undiscovered particles (like the Higgs boson) below $\Lambda\sim $few TeV.
Then, at the LEP energy, the only particles (unphysical) related to the EWSBM 
are the would-be Goldstone bosons (GBs). The system of the GBs
and the electroweak gauge bosons can be generally described by the electroweak 
chiral Lagrangian (EWCL) \cite{AW} which can be regarded as the low energy 
effective Lagarangian of the fundamental theory of EWSBM, and can be expanded 
according to the powers of $p^2/\Lambda^2$, 
\begin{eqnarray}                   
{\cal L}= {\cal L}^{(2)}+{\cal L}^{(4)}+\cdots\,,
\label{L}
\end{eqnarray}
where ${\cal L}^{(2)}$ and ${\cal L}^{(4)}$ are terms of $O(p^2/\Lambda^2)$ 
and $O(p^4/\Lambda^4)$, respectively. Actually, the $Z$-pole observables are 
not sensitive to ${\cal L}^{(4)}$, so that the authors mainly considered 
${\cal L}^{(2)}$ in which there are two terms related to $S$ and $T$. 
The authors made a model-independent analysis with $S$ and $T$ taken as 
two unknowns which, together with the QCD coupling constant $\alpha_s$, are 
adjusted to make the best fit of the EWCL (\ref{L}) to the $Z$-pole data. 
Their result shows that with the best fit values
\begin{eqnarray}                    
&&S=-0.13\pm 0.10\,,~~~~T=0.13\pm 0.11\,,~~~~\alpha_s(M_Z)=0.119\pm 0.003\,,
\end{eqnarray}
the $Z$-pole data can be well fitted. The best fit value of
$\alpha_s(M_Z)$ is almost the same as the world averaged value, so that
the result is very reasonable. This result means that {\it the $Z$-pole data
can be well fitted even without a Higgs boson below the scale $\Lambda$}.

We see from the above analyses that the hints of the Higgs mass from the
best fit to the LEP data are quite different in different approaches
with or without new physics contributions. We can conclude that {\it
the LEP/SLD precision $Z$-pole data do not necessarily imply the
existence of a light Higgs boson}, so that the probe of the EWSB mechanism
should be proceeded in a wide scope considering both the case of existing a 
light Higgs boson and the case without a Higgs boson below the scale of
TeV. Note that the width of the Higgs boson is proportional to $m_H^3$,
so that a light Higgs boson will look as a narrow resonance which is
easy to detect. If a Higgs boson is so heavy that its width is comparable to 
its mass, it will not show up as a clear resonance, and the detection is hard. 
In this case or there is no Higgs resonance below 1 TeV, other method of 
probing the EWSB mechanism should be developed. We shall deal with this problem
in Sec. 4.

\subsection{Searching for the Higgs Boson at High Energy Colliders}

 Searching for the Higgs boson is the first important task at the
future high energy colliders. Here we briefly review various ways of searching
for a light SM Higgs boson at high energy colliders.  
\newpage
\null\noindent{\bf i, LEP2} \cite {LEP2}

At the LEP2 energy, the dominant production mechanism for the EWSM Higgs 
boson is the Higgs-strahlung process
\begin{eqnarray}                      
e^+e^-\to Z^*\to Z~H\,,
\label{Higgs-strahlung}
\end{eqnarray}
in which the Higgs boson is emitted from a virtual Z boson. 
The latest experiments were the 1999 run of LEP2 at $\sqrt{s}=192$ GeV and
$\sqrt{s}=202$ GeV in which no evidence of the Higgs boson was found.
This leads to the $95\%$ C.L. lower bound on $m_H$ \cite{LEP-m_H}
\begin{eqnarray}                   
m_H>107.7~ {\rm GeV}\,.
\end{eqnarray}

\null\noindent{\bf ii, Upgraded Tevatron}

It has been shown that at the upgraded Tevatron Run 2, the most
promising process for the search for the Higgs boson is
\begin{eqnarray}                    
p\bar p\to WH\,,~~~~~~~~~~p\bar p\to ZH\,,
\end{eqnarray}
with the tagging channel $H\to b\bar b$. Together with the tagging mode
$H\to \tau^+\tau^-$, the searching ability can be up to
$m-H<130$ GeV \cite{MK}. This is just not enough to cover the interesting 
theoretical upper limit of the lightest MSSM Higgs boson $h$, $m_h<130\pm 5$ 
GeV \cite{HHW}. Recently an interesting investigation was made in Ref. 13 
showing that the EWSM Higgs boson in the mass region 
$135~{\rm GeV}\leq m_H\leq 180~{\rm GeV}$ is able to be detected at the 
upgraded Tevatron with the $\sqrt{s}=2$ TeV and 
an integrated luminosity of 30 fb$^{-1}$/yr (Run 3 of the Tevatron) via the 
process \cite{HZ}
\begin{eqnarray}                    
p\bar{p}\to gg\to H\to W^*W^*\to l\nu jj\,,~~l\bar{\nu}\bar{l}\nu\,.
\end{eqnarray}
Therefore the upgraded Tevatron will be the next collider of Higgs searching 
after LEP2. Of course, due to the low luminosity, it will take years to 
accumulate enough events to draw a firm conclusion.

\null\noindent{\bf iii, LHC}

At the LHC, because of the hadronic backgrounds, searching for the
Higgs boson of mass $m_H>140$ GeV and $m_H<140$ GeV are quite
different. In the following, we review these two kinds of searches separately.
\begin{itemize}
\item {\bf $m_H>140$ GeV}

~~~~In this case the following {\it gold plated channel} is available 
\cite{LHCW}
\begin{eqnarray}                             
&&pp\to HX\to ZZ(Z^*)X\to l^+l^-l^+l^-X~({\rm or}~l^+l^-\nu\bar{\nu}X)\,,
\nonumber\\
&&pp\to HX\to WW(W^*)X\to l^+l^-\nu\bar{\nu}X)\,,
\end{eqnarray}
in which the four-lepton final state is very clear with rather small 
backgrounds. 
Theoretical study shows that the resonance behavior can be
clearly seen when $m_H<800$~GeV. When $m_H\geq 800$ GeV,
the width of the Higgs boson will be comparable to the mass, and
the Higgs boson can hardly be seen as a resonance. Searching
for such a heavy Higgs boson, as well as probing the EWSBM
when there is no Higgs resonance below 1 TeV, will be reviewed in Sec. 4.
\item {\bf $M_Z<m_H<140$ GeV}

~~~~If the Higgs mass is in the intermediate range $M_Z<m_H<140~$GeV, the 
above detection is not possible since the branching ratio of the four-lepton 
channel drops very rapidly as $m_H<140~$GeV.
Detection of such an intermediate-mass SM Higgs boson is much more difficult. 
Fortunately, the $H\to \gamma\gamma$ branching ratio has its maximal
value in this $m_H$ range. Thus the best way is to detect the $\gamma\gamma$ 
final state for the Higgs boson. Recently, it is shown that the EWSM
Higgs boson in the mass range of 100 GeV$\--$150 GeV can be detected at
the LHC via
\begin{eqnarray}                   
pp\to H(\gamma\gamma)+jet
\end{eqnarray}
if a transverse-momentum cut of 2 GeV on the tracks is made for reducing the 
background \cite{KZ}. 

~~~~To find channels with better signal to background ratio, people suggested 
the following associate productions of $H$ \cite{A-P}.
\begin{eqnarray}                    
pp\to WHX\to l\bar{\nu}\gamma\gamma X\,,~~~~~~
pp\to t\bar{t}HX\to l\bar{\nu}\gamma\gamma X\,.
\end{eqnarray}
The signal and backgrounds of the $WH$ associate production channel have 
been calculated in Ref. 15 which shows that the backgrounds are smaller 
than the signal even for a mild photon detector with a $3\%$ $\gamma\gamma$ 
resolution. The inclusive search for the $t\bar{t}H$ associate production 
suffers from a further large background from 
$pp\to W(\to l\bar{\nu})\gamma\gamma(n-jet),~(n=1,\cdots,4)$ \cite{ZK}, and 
the search is possible only when the $\gamma\gamma$ resolution of the photon 
detector is of the level of $1\%$ \cite{ZK}. The photon detectors of
ATLAS and CMS at the LHC are just of this level. Actually, if the 
jets are also detected, the background can be effectively reduced with suitable
choice of the jets, and such detection is possible even for the mild photon 
detector with $3\%$ $\gamma\gamma$ resolution \cite{ZK}.

~~~~Recently, the $b$-tagging efficiency is much improved. Tagging a
light Higgs boson (with large enough $B(H\to b\bar{b})$)
via the $H\to b\bar{b}$ mode with a detectable signal to background
ratio is already possible at LHC. The number of events will be larger
than that in the $H\to \gamma\gamma$ tagging mode.

~~~~In summary, a Higgs boson with mass $m_H<800$ GeV can definitely be 
detected as a resonance at the LHC.
\end{itemize}

\null\noindent{\bf iv, LC}

The advantage of searching for the Higgs boson at the LC is the smallness
of the hadronic backgrounds. Then, the $H\to b\bar{b}$ mode can be
taken as the main tagging mode to have larger number of events.

At the LC, the Higgs boson can be produced either by the
Higgs-strahlung process (\ref{Higgs-strahlung}) or by $WW$ and $ZZ$ fusions
\begin{eqnarray}                      
e^+e^-\to \nu\bar{\nu}(WW)\to \nu\bar{\nu}H\,,~~~~~~
e^+e^-\to e^+e^-(ZZ)\to e^+e^-H\,.
\end{eqnarray}
The cross sections for the Higgs-strahlung and $WW$ fusion processes are
$\sigma\sim 1/s$ and $\sigma\sim (\ln\frac{s}{M_W})/M^2_W$, 
respectively.
So that 
the Higgs-strahlung process is important at $\sqrt{s}\leq 500$
GeV, while the $WW$ fusion process is important
at $\sqrt{s}>500$ GeV. With the
$H\to b\bar{b}$ tagging mode, several thousands of events can be produced
for the envisaged luminosities \cite{LCPhys}.

Furthermore, by means of laser back-scattering, $\gamma\gamma$ and 
$e\gamma$ colliders can be constructed based on the LC. It has been shown 
recently that the $s$-channel Higgs production rate at the photon collider 
will be about an order of magnitude larger than the production rate in the 
Higgs-strahlung process at the LC \cite{Telnov}.

\subsection{Testing Higgs Boson Interactions}
 
If a light Higgs resonance is found from the above searches, it is {\it not 
the end of the story}. It is needed to test whether it is the EWSM Higgs or 
something else .
This can be done by examining its interactions. We know the
self-interactions of the SM Higgs boson contain the following trilinear
and quartic terms
\begin{eqnarray}                     
\frac{1}{8}\frac{m_H^2}{v^2}\bigg[4vH^3+H^4\bigg]\,,
\end{eqnarray}
where $v=246$ GeV is the VEV of the Higgs field.
For detecting the trilinear interaction, it is possible to look at the
double Higgs-boson productions $pp\to HHX$ at the LHC and
$e^+e^-\to HHZ,~~HH\bar{\nu}_e\nu_e$ at the LC. It has been shown that
the detection at the LHC is almost impossible due to the large background 
\cite{LCPhys,KZ}, while the detection at the LC is possible at the
C.M. energy $E=1.6$ TeV requiring a very large integrated luminosity,
$\int{\cal L}dt=1000$ fb$^{-1}$. Therefore the detection is not easy.
The signals of the quartic interactions are so small that it is hard to
detect. 

Since the top quark has the largest Yukawa coupling to the EWSM Higgs boson,
it is possible to detect the Higgs Yukawa coupling via the process
\begin{eqnarray}                             
e^+e^-\to t\bar{t}H\,,
\label{ttH}
\end{eqnarray}
which can test the $Ht\bar{t}$ Yukawa coupling and see whether the discovered
Higgs boson is really the one responsible for the top quark mass.
This detection has been studied in Refs.\cite{Dawson,ttH}.

\section{Strongly Interacting Electroweak Symmetry Breaking Mechanism}

Introducing elementary Higgs field is the simplest but not unique EWSBM.
The way of completely avoiding triviality and unnaturalness is to abandon 
elementary scalar fields and introducing new strong interactions causing 
certain fermion condensates to break the electroweak gauge symmetry. This idea 
is similar to those in the theory of superconductivity and chiral symmetry 
breaking in QCD. The simplest model realizing this idea is the original 
QCD-like technicolor (TC) model. However, such a simple model 
predicts a too large value of $S$ and is already ruled out by the LEP data. A 
series of improved models have been proposed to overcome the shortcomings of 
the simplest model. In the following, we briefly review two of the recently 
proposed models.

\null\noindent{\bf i, Topcolor-Assisted Technicolor Models}

This model combines the technicolor and the top-condensate ideas
\cite{TC2}. It is assumed in this model that at the energy scale
$\Lambda\sim 1$ TeV, there is a {\it topcolor} theory with the gauge 
group $SU(3)_1\times U(1)_{Y1}\times SU(3)_2\times U(1)_{Y2}\times SU(2)_L$
in which $SU(3)_1\times U(1)_{Y1}$ preferentially couples to the third-family
fermions and $SU(3)_2\times U(1)_{Y2}$ preferentially couples to the first- 
and second-family fermions.
It is assumed that there is also a TC sector which is the main part in the
EWSBM and will break the topcolor gauge group into $SU(3)_{QCD}$ and $U(1)_Y$ 
at the scale $\Lambda$.
The $SU(3)_1\times U(1)_{Y1}$ couplings are assumed to be much stronger
than those of $SU(3)_2\times U(1)_{Y2}$. The strong $SU(3)_1\times U(1)_{Y1}$
interactions will form top quark condensate $\langle t\bar{t}\rangle$ but not 
bottom quark condensate from the simultaneous effects of the $SU(3)_1$ and
$U(1)_{Y1}$ interactions. The TC dynamics gives rise to the
masses of the $u,~d,~s,~c,~$ and $b$ quarks and a small portion of the
top quark mass, while the main part of the top quark mass comes from
the topcolor dynamics causing the top quark condensate just like the
constituent quarks acquiring their large dynamical masses from the dynamics
causing the quark condensates in QCD. In this prescription, the
TC dynamics does not cause a large oblique correction
parameter $T$ even the mass difference $m_t-m_b$ is so large. 
Improvement of this kind of model is still in progress. 

This kind of model contains various pseudo-Goldstone bosons (PGBs)
including {\it technipions} in the techicolor sector and an isospin
triplet {\it top-pions} with masses in a few hundred GeV range. It has been 
shown that 
the LEP/SLD data of $R_b$ put constraint on the top-pion mass \cite{TC2_Rb-2}. 
These light particles characterizing the phenomenology of the model.

\null\noindent{\bf ii, Top Quark Seesaw Theory}

Recently, a new promising theory of strongly interacting 
EWSB related to the top quark condenstate called {\it top quark seesaw theory} 
was proposed in Ref. 22. The gauge group in this theory is \cite{topseesaw}
\begin{eqnarray}                    
G\times G_{tc}\times SU(2)_W\times U(1)_Y\,,
\label{gauge}
\end{eqnarray}
where $G_{tc}$ is the topcolor gauge group (for instance, 
$SU(3)_1\times SU(3)_2$ or even larger), $G$ is a gauge group for new
strong interactions which breaks $G_{tc}$ into $SU(3)_{QCD}\times U(1)_Y$ at a
scale $\Lambda$. Instead of introducing techniquarks, certain 
$SU(2)_W$-singlet quarks, $\chi,\cdots$, with topcolor interactions and
specially assigned $U(1)_Y$ quantum numbers are introduced in 
this theory. For instance, the simplest model can be constructed by
assigning the left-handed third family quark-field $\psi_L$, 
the right-handed top quark $t_R$, and an $SU(2)_W$-singlet quark $\chi$ 
in the following representations of $SU(3)_1\times SU(3)_2\times SU(2)_W
\times U(1)_Y$
\begin{eqnarray}              
\psi_L:~({\bf 3,~1,~2},~+1/3)\,,~~~~\chi_R:~({\bf 3,~1,~1,},~+4/3)\,,~~~~
t_R,~\chi_L:~({\bf 1,~3,~1},~+4/3)\,.
\label{QN}
\end{eqnarray}
Topcolor will cause the following $t$ ($b$) and $\chi$ bound state
scalar field
\begin{eqnarray}             
\varphi=\pmatrix{\overline{\chi_R}~t_L\cr \overline{\chi_R}~b_L}
\label{cHiggs}
\end{eqnarray}
which behaves like a Higgs doublet whose VEV breaks the electroweak symmetry.
Furthermore, the VEV of $\varphi$ will cause a dynamical mass 
$m_{t\chi}\sim 600$ GeV, and the dynamics in this theory causes a
seesaw mechanism for the mass terms in the $\chi-t$ sector which leads
to the following top quark mass
\begin{eqnarray}                    
m_t\approx m_{t \chi}\frac{\mu_{\chi t}}{\mu_{\chi\chi}}\,,
\label{mt}
\end{eqnarray}
where $\frac{\mu_{\chi t}}{\mu_{\chi\chi}}$ is determined by the dynamics
and can yield the desired top quark mass.

This theory has several advantages. (a) In this model, one of the 
particles responsible for the EWSBM is just the known top quark, and the 
$SU(2)_W$-doublet nature of the Higgs filed just comes from the 
same nature of the third family quarks. (b) The new quark $\chi$ introduced in 
this theory is $SU(2)_W$-singlet so that there is no large custodial symmetry 
violation causing a too large $T$.
(c) The problem of predicting a too large $S$ in technicolor theories due to 
introducing many technifermion-doublets does not exist in the present theory
since there is only one top quark condensate. (d) Unlike
the original top quark condensate model which leads to a too large top
quark mass, the present theory can give rise to the desired top quark mass 
via the seesaw mechanism.

There can be various ways of building realistic models in this theory.
Very recently, two realistic models which can fit all the precision 
electroweak data have been built in Ref. 22. We briefly review these two
models.

The first model is a one-Higgs-doublet model with the composite Higgs field
$\varphi$ defined in (\ref{cHiggs}). The precision data can be fitted with
\cite{topseesaw}
\begin{eqnarray}                  
m_H\sim 0.5\--1~{\rm TeV}
\end{eqnarray}
corresponding to $~m_{\chi}\sim 5\--8~$ TeV. The lower limit of $m_H$ is
$~m_H|_{min}=159$ GeV corresponding to $~m_{\chi}\to \infty$.

The second model is a two-Higgs doublet model. In addition to the
$SU(2)_W$-singlet quark $\chi$ introduced in (\ref{QN}), another 
$SU(2)_W$-singlet quark in the representation
\begin{eqnarray}               
\omega_R:~({\bf 3,~1,~1},~-2/3)\,,~~~~~~~~
b_R,~\omega_L:~({\bf 1,~3,~1},!-2/3)
\end{eqnarray}
is intrduced. Then, $\chi_R$ and $\omega-R$ can form a doublet
\begin{eqnarray}               
\lambda_R=\pmatrix{\chi_R\cr\omega_R}\,,
\end{eqnarray}
and two composite Higgs doublets can be formed by the composite object
\begin{eqnarray}               
\overline{\lambda_R}\psi_L\,.
\end{eqnarray}
It contains the three would-be Goldstone bosons and five Higgs bosons:
two CP even neutral scalar Higgs fields $h^0$ and $H^0$, one CP odd 
pseudoscalar Higgs field $A^0$, and a pair of charged Higgs $H^\pm$.
The precision data can be fitted with \cite{topseesaw}
\begin{eqnarray}               
m_A\sim 100~{\rm GeV}\,,~~~~~~~~~~~~~~~m_{h^0}\sim m_{H^0}\sim
m_{H^\pm}\sim 800~{\rm GeV}
\end{eqnarray}
corresponding to $m_\chi\sim 3\--5$ TeV and $m_\omega\sim 12$ TeV.

These results are obtained from quite complicated arrangements of the
gauge group $G$ \cite{topseesaw}, and there may exist some extra
scalar (pseudocalar) bound states in the theory\cite{topseesaw}.

Due to the nonperturbative nature of the strong interaction dynamics,
it is hard to make precision predictions from the strongly interacting
EWSM. However, some models contain extra heavy gauge bosons below 1 TeV, and 
most of the models contain certain model-dependent PGBs with masses in the 
region of few hundred GeV. Their effects can be experimentally tested.

Direct productions of PGBs
have been extensively studied in the literature \cite{PGBprod,HY}. It is 
shown that the detection are possible but not all easy.

Since the top quark couples to the EWSB sector strongly due to its large mass, 
a feasible way of testing the strongly interacting EWSBM is to test the
extra gauge boson and PGB effects in top quark productions at high energy 
colliders. This kind of study has been carried out in various papers 
\cite{ttprod}
The conclusions of these studies are that not only the PGB 
effects can be detected, but also different models can be experimentally
distinguished (also can be distinguished from the MSSM) by measuring the 
production cross sections and the invariant mass distributions \cite{TC2-tt}.

\section{Model-Independent Probe of Elwctroweak Symmetry Breaking
Mechanism}

We have seen that there are various kinds of EWSBMs proposed. 
We do not know whether the actual EWSBM in the nature looks like one of them 
or not. Therefore, only testing the proposed models seems to be not enough, 
and certain model-independent probe of the EWSBM is needed. Since the
scale of new physics is likely to be several TeV, electroweak physics at energy
$E\leq 1$ TeV can be effectively described by the {\it electroweak
effective Lagrangian} in which composite fields are approximately
described by effective local fields. The electroweak effective Lagrangian is a 
{\it general description} (including all kinds of models) which contains 
certain yet unknown coefficients whose values are, in principle, 
determined by the underlying dynamics. Different EWSBMs give rise to 
different sets of coefficients. The model-independent probe is to investigate 
through what processes and to what precision we can measure these coefficients 
in the experiments. 
From the experimental point of view, the most challenging case of probing
the EWSBM is that there is no light scalar resonance found
below 1 TeV. We shall take this case as the example in this review.
Effective Lagrangian including a light Higgs boson has also been studied in the
literature \cite{Buchmuller}. In the case we are considering, the
effective Lagrangian is the so called electroweak chiral Lagrangian (EWCL)
which is a Lagrangian for the would-be Goldstone bosons $\pi^a$
in the nonlinear realization $U=e^{i\tau^a\pi^a/f_\pi}$ with
electroweak interactions. The bosonic sector of which, up 
to the $p^4$-order, reads \cite{AW,global}
\begin{eqnarray}                   
&&{\cal L}_{\rm eff}={\cal L}_{\rm G}+{\cal L}_{\rm S}\,,
\label{CL1}
\end{eqnarray}
where ${\cal L}_{\rm G}$ is the weak gauge boson kinetic energy term, and
\begin{eqnarray}                  
{\cal L}_{\rm S}= {\cal L}^{(2)}+{\cal L}^{(2)\prime}+
             \displaystyle\sum_{n=1}^{14} {\cal L}_n ~~,
\label{CL2}		   
\end{eqnarray}
with 
\vspace{0.2cm}
\begin{eqnarray}                  
&&{\cal L}^{(2)}=\frac{f_\pi^2}{4}{\rm Tr}[(D_{\mu}U)^\dagger(D^{\mu}U)]~~,
~~~~~~~~~~~~~~
{\cal L}^{(2)\prime} =\ell_0 (\frac{f_\pi}{\Lambda})^2~\frac{f_\pi^2}{4}
               [ {\rm Tr}({\cal T}{\cal V}_{\mu})]^2 ~~,\nonumber\\
&&{\cal L}_1 = \ell_1 (\frac{f_\pi}{\Lambda})^2~ \frac{gg^\prime}{2}
B_{\mu\nu} {\rm Tr}({\cal T}{\bf W^{\mu\nu}}) ~~,~~~~~~~~
{\cal L}_2 = \ell_2 (\frac{f_\pi}{\Lambda})^2 ~\frac{ig^{\prime}}{2}
B_{\mu\nu} {\rm Tr}({\cal T}[{\cal V}^\mu,{\cal V}^\nu ]) ~~,\nonumber\\
&&{\cal L}_3 = \ell_3 (\frac{f_\pi}{\Lambda})^2 ~ig
{\rm Tr}({\bf W}_{\mu\nu}[{\cal V}^\mu,{\cal V}^{\nu} ]) ~~,~~~~~~~~
{\cal L}_4 = \ell_4 (\frac{f_\pi}{\Lambda})^2 
                     [{\rm Tr}({\cal V}_{\mu}{\cal V}_\nu )]^2 ~~,\nonumber\\
&&{\cal L}_5 = \ell_5 (\frac{f_\pi}{\Lambda})^2 
                     [{\rm Tr}({\cal V}_{\mu}{\cal V}^\mu )]^2 ~~,
				 ~~~~~~~~~~~~~~~~~~
{\cal L}_6 = \ell_6 (\frac{f_\pi}{\Lambda})^2 
[{\rm Tr}({\cal V}_{\mu}{\cal V}_\nu )]
{\rm Tr}({\cal T}{\cal V}^\mu){\rm Tr}({\cal T}{\cal V}^\nu) ~~,\nonumber\\
&&{\cal L}_7 = \ell_7 (\frac{f_\pi}{\Lambda})^2 
[{\rm Tr}({\cal V}_\mu{\cal V}^\mu )]
{\rm Tr}({\cal T}{\cal V}_\nu){\rm Tr}({\cal T}{\cal V}^\nu) ~~,~~~~
{\cal L}_8 = \ell_8 (\frac{f_\pi}{\Lambda})^2~\frac{g^2}{4} 
[{\rm Tr}({\cal T}{\bf W}_{\mu\nu} )]^2  ~~,\nonumber\\
&&{\cal L}_9 = \ell_9 (\frac{f_\pi}{\Lambda})^2 ~\frac{ig}{2}
{\rm Tr}({\cal T}{\bf W}_{\mu\nu}){\rm Tr}
        ({\cal T}[{\cal V}^\mu,{\cal V}^\nu ]) ~~,~~~~
{\cal L}_{10} = \ell_{10} (\frac{f_\pi}{\Lambda})^2\frac{1}{2}
[{\rm Tr}({\cal T}{\cal V}^\mu){\rm Tr}({\cal T}{\cal V}^{\nu})]^2 ~~,\nonumber\\
&&{\cal L}_{11} = \ell_{11} (\frac{f_\pi}{\Lambda})^2 
~g\epsilon^{\mu\nu\rho\lambda}
{\rm Tr}({\cal T}{\cal V}_{\mu}){\rm Tr}
({\cal V}_\nu {\bf W}_{\rho\lambda}) ~~, ~~~~
{\cal L}_{12} = \ell_{12}(\frac{f_\pi}{\Lambda})^2 ~2g
                    {\rm Tr}({\cal T}{\cal V}_{\mu}){\rm Tr}
                  ({\cal V}_\nu {\bf W}^{\mu\nu}) ~~,\nonumber\\
&&{\cal L}_{13} = \ell_{13}(\frac{f_\pi}{\Lambda})^2~ 
      \frac{gg^\prime}{4}\epsilon^{\mu\nu\rho\lambda}
      B_{\mu\nu} {\rm Tr}({\cal T}{\bf W}_{\rho\lambda}) ~~,\nonumber\\
&&{\cal L}_{14} = \ell_{14} (\frac{f_\pi}{\Lambda})^2~\frac{g^2}{8} 
\epsilon^{\mu\nu\rho\lambda}{\rm Tr}({\cal T}{\bf W}_{\mu\nu})
{\rm Tr}({\cal T}{\bf W}_{\rho\lambda})~~,
\label{CL3}
\end{eqnarray}
\vspace{0.2cm}
in which $D_{\mu}U =\partial_{\mu}U + ig{\bf W}_{\mu}U 
-ig^{\prime}U{\bf B}_{\mu}~,~~~{\cal V}_{\mu}\equiv (D_{\mu}U)U^\dagger~$, 
~~and $~~{\cal T}\equiv U\tau_3 U^{\dagger}$. 

The coefficients $\ell$'s reflect the strengths of the $\pi^a$
interactions, i.e. the EWSBM. $\ell_1,~\ell_0$ and $\ell_8$ are
related to the oblique correction parameters $S,~T$ and $U$,
respectively; $\ell_2,~\ell_3,~\ell_9$ are related to the 
triple-gauge-couplings;
${\cal L}_{12},~{\cal L}_{13}$ and ${\cal L}_{14}$ are CP-violating.
The task now is to find out experimental processes to measure the yet 
undetermined $\ell$'s. 

Note that $\pi^a$ are not physical particles, so that they are not 
experimentally observable. However, due to the 
Higgs mechanism, the degrees of freedom of $\pi^a$ are related to the 
longitudinal components of the weak bosons $V^a_L$ ($W^\pm_L,~Z^0_L$) which 
are experimentally observable. Thus the $\ell$'s are able to be
measured via $V^a_L$-processes. So that we need to know the quantitative 
relation between the $V^a_L$-amplitude (related to the experimental data) and 
the GB-amplitude (reflecting the EWSB mechanism), which is the so-called 
{\it equivalence theorem} (ET). ET has been studied by many papers, and the 
final precise formulation of the ET and its rigorous proof are given in 
Refs.\cite{et}. The precise formulation of the
ET is
\begin{eqnarray}                       
T[V^{a_1}_L,V^{a_2}_L,\cdots]                                   
= C\cdot T[-i\pi^{a_1},\i\pi^{a_2},\cdots]+B ~~,
\label{ET}
\end{eqnarray}
with
\begin{eqnarray}                       
&&E_j \sim k_j  \gg  M_W , ~~~~~(~ j=1,2,\cdots ,n ~)~~,\nonumber\\
&&C\cdot T[-i\pi^{a_1},-i\pi^{a_2},\cdots]\gg B ~~,
\end{eqnarray}
where $~T[V^{a_1}_L,V^{a_2}_L,\cdots]~$ and 
$~T[-i\pi^{a_1},-i\pi^{a_2},\cdots]~$
are, respectively, the $V^a_L$-amplitude and the $\pi^a$-amplitude$, E_j$ is 
the energy of the $j$-th external line, $C$ is a gauge and
renormalization scheme dependent constant factor, and $B$ is a 
process-dependent function of the energy $E$. 
By taking special convenient renaormalization scheme, the constant $C$ can be
simplified to $C=1$ \cite{et}. In the EWCL theory, the
$B$-term may not be small even when the center-of-mass energy $E\gg M_W$, and 
{\it it is not sensitive to the EWSB mechanism}. Therefore the $B$-term serves 
as an {\it intrinsic background} when probing $\pi^a$-amplitude
via the $V^a_L$-amplitude in (\ref{ET}). Only when 
$~|B|\ll |C\cdot T[-i\pi^{a_1},-i\pi^{a_2},\cdots]|~$ the probe can be
{\it sensitive}. In Ref. 29, a new power counting rule for 
semi-quantitatively estimating the amplitudes in the EWCL theory was proposed, 
and with which a systematic analysis on the sensitivities of probing the EWSB
mechanism via the $V^a_L$ processes were given. The results are
summarized in Table 1.

We see that the coefficients $\ell$'s can be experimentally determined 
via various $V^a_L$ processes at various phases of the LHC and the LC 
(including the $e\gamma$ collider) complementarily. Without the LC, the LHC 
itself is not enough for determining all the coefficients. Quantitative
calculations on the determination of the quartic-$V^a_L$-couplings

\vspace{0.2cm}
\tabcolsep 1pt
\null\noindent{\footnotesize {\bf TABLE I.}  Probing the EWSB Sector at High 
Energy Colliders: A Global Classification for the NLO Bosonic Operators. 
Quoted from Ref. 28. 
\\ \\
(~Notations: ~$\surd =~$Leading contributions, 
$~\triangle =~$Sub-leading contributions,~ 
and ~$\bot =~$Low-energy contributions.~
~Notes:~ $^{\dagger}$Here, $~{\cal L}_{13}$ or $~{\cal L}_{14}~$ 
does not contribute at $~O(1/\Lambda^2)~$.   ~~ $^\ddagger$At LHC($14$),  
$W^+W^+\to W^+W^+$ should also be included.~)}
{\scriptsize
\vspace{-0.6cm}

\begin{tabular}{||c||c|c|c|c|c|c|c|c|c|c||c||c||} 
\hline\hline
& & & & & & & & & & & & \\
 Operators 
& $ {\cal L}^{(2)\prime} $ 
& $ {\cal L}_{1,13} $ 
& $ {\cal L}_2 $
& $ {\cal L}_3 $
& $ {\cal L}_{4,5} $
& $ {\cal L}_{6,7} $ 
& $ {\cal L}_{8,14} $ 
& $ {\cal L}_{9} $
& $ {\cal L}_{10} $
& $ {\cal L}_{11,12} $
& $T_1~\parallel  ~B$ 
& Processes \\
& & & & & & & & & & & & \\
\hline\hline
 LEP-I (S,T,U) 
& $\bot$ 
& $\bot~^\dagger$
&  
& 
& 
& 
& $\bot~^\dagger$
& 
&
&
& $g^4\frac{f^2_\pi}{\Lambda^2}$ 
& $e^-e^+\to Z \to f\bar{f}$\\ 
\hline
  LEP-II
& $\bot$ 
& $\bot$  
& $\bot$  
& $\bot$  
&  
& 
& $\bot$  
& $\bot$ 
&
& $\bot$  
& $g^4\frac{f^2_\pi}{\Lambda^2}$
& $e^-e^+ \to W^-W^+$\\
\hline
  LC($0.5$)/LHC($14$)
& 
& 
& $\surd$
& $\surd$
& 
& 
& 
& $\surd$
&
& 
& $g^2\frac{E^2}{\Lambda^2} \parallel g^2\frac{M_W^2}{E^2}$
& $f \bar f\to W^-W^+ /(LL)$\\  
& 
& $\triangle$
& $\triangle$
& $\triangle$
& 
& 
& $\triangle$
& $\triangle$
&
& $\triangle$
& $g^3\frac{Ef_\pi}{\Lambda^2} \parallel g^2\frac{M_W}{E} $ 
& $f \bar f\to W^-W^+/(LT) $\\  
\hline
& 
& 
& 
& $\surd$
& $\surd$
& $\surd$
& 
& $\surd$
&
& $\surd$
& $g^2\frac{1}{f_\pi}\frac{E^2}{\Lambda^2}
  \| g^3\frac{M_W}{E^2} $
& $f \bar f\to W^-W^+Z /(LLL) $\\
& 
& $\triangle$ 
& $\triangle$
& $\triangle$
& $\triangle$
& $\triangle$
& $\triangle$
& $\triangle$
&
& $\triangle$
& $g^3\frac{E}{\Lambda^2}\parallel g^3\frac{M_W^2}{E^3}$ 
& $f \bar f\to W^- W^+ Z /(LLT)  $\\
& 
& 
&  
& $\surd$
& $\surd$
& $\surd$
& 
& 
& $\surd$
& 
& $g^2\frac{1}{f_\pi}\frac{E^2}{\Lambda^2}\parallel 
  g^3\frac{M_W}{\Lambda^2}$
& $f \bar f \to ZZZ /(LLL) $\\
& 
& 
& 
& 
& $\triangle$
& $\triangle$
& 
& 
& $\triangle$
& 
& $g^3\frac{E}{\Lambda^2}\parallel
   g^3\frac{f_\pi}{\Lambda^2}\frac{M_W}{E}$ 
& $f \bar f \to ZZZ  /(LLT)  $\\
 ~LC($1.5$)/LHC($14$)~ 
& 
& 
& 
& 
& $\surd$
& 
& 
&
& 
& 
& $\frac{E^2}{f_\pi^2}\frac{E^2}{\Lambda^2}\parallel g^2$ 
& $W^-W^\pm \to W^-W^\pm /(LLLL)~^\ddagger$\\
&
& 
& 
& $\triangle$
& $\triangle$
& 
& 
& $\triangle$
&
& $\triangle$
& $g\frac{E}{f_\pi}\frac{E^2}{\Lambda^2}\parallel g^2\frac{M_W}{E}$ 
& $W^-W^\pm\to W^-W^\pm /(LLLT)~^\ddagger$ \\
& 
& 
& 
& 
& $\surd$
& $\surd$
&
& 
&
& 
& $\frac{E^2}{f_\pi^2}\frac{E^2}{\Lambda^2}\parallel g^2 $
& $W^-W^+ \to ZZ ~\&~{\rm perm.}/(LLLL)$ \\
&
& 
& $\triangle$
& $\triangle$
& $\triangle$
& $\triangle$
& 
& $\triangle$
&
& $\triangle$
& $g\frac{E}{f_\pi}\frac{E^2}{\Lambda^2}\parallel g^2\frac{M_W}{E}$ 
& $W^-W^+ \to ZZ ~\&~{\rm perm.} /(LLLT)$ \\
& 
& 
& 
& 
& $\surd$
& $\surd$
& 
& 
& $\surd$ 
&  
& $\frac{E^2}{f_\pi^2}\frac{E^2}{\Lambda^2}\parallel
   g^2\frac{E^2}{\Lambda^2} $
& $ZZ\to ZZ /(LLLL) $\\
&
& 
& 
& $\triangle$
& $\triangle$
& $\triangle$
&  
&
& $\triangle$
&
& $g\frac{E}{f_\pi}\frac{E^2}{\Lambda^2}\parallel
  g^2\frac{M_WE}{\Lambda^2}$ 
& $ZZ\to ZZ /(LLLT) $\\
\hline
& 
& 
& 
& $\surd$
& 
& 
& 
&
& 
& $\surd$
& $g^2\frac{E^2}{\Lambda^2} \parallel  g^2\frac{M^2_W}{E^2}$
& $q\bar{q'}\to W^\pm Z /(LL) $\\
& 
& $\triangle$
& $\triangle$
& $\triangle$
& 
& 
& $\triangle$
& $\triangle$
&
& $\triangle$
& $g^3\frac{Ef_\pi}{\Lambda^2}\parallel g^2\frac{M_W}{E}$ 
& $q\bar{q'}\to W^\pm Z /(LT) $\\
 LHC($14$)
& 
& 
& 
& $\surd$
& $\surd$
& 
& 
& $\surd$
&
& $\surd$
& $g^2\frac{1}{f_\pi}\frac{E^2}{\Lambda^2}\parallel g^3\frac{M_W}{E^2}$
& $q \bar{q'}\to W^-W^+W^\pm /(LLL) $\\
& 
& 
& $\triangle$
& $\triangle$
& $\triangle$
&
& $\triangle$
& $\triangle$
&
& $\triangle$
& $g^3\frac{E}{\Lambda^2}\parallel g^3\frac{M_W^2}{E^3}$ 
& $q \bar{q'} \to W^- W^+W^\pm  /(LLT)  $\\
& 
& 
& 
& $\surd$
& $\surd$
& $\surd$
& 
& 
&
& $\surd$
& $g^2\frac{1}{f_\pi}\frac{E^2}{\Lambda^2}\parallel g^3\frac{M_W}{E^2}$
& $q \bar{q'}\to W^\pm ZZ /(LLL) $\\
& 
& $\triangle$
& $\triangle$
& $\triangle$
& $\triangle$
& $\triangle$
& $\triangle$
& $\triangle$
&
& $\triangle$
& $g^3\frac{E}{\Lambda^2}\parallel  g^3\frac{M_W^2}{E^3}$ 
& $q \bar{q'} \to W^\pm ZZ  /(LLT)  $\\
\hline
 LC($e^-\gamma$)
& 
& $\surd$
& $\surd$
& $\surd$
& 
& 
& $\surd$
& $\surd$
&
& $\surd$
& $eg^2\frac{E}{\Lambda^2}\parallel eg^2\frac{M_W^2}{E^3}$ 
& $e^-\gamma \to \nu_e W^-Z,e^-WW /(LL)$\\
\hline
& 
& $\surd$
& $\surd$
& $\surd$
& 
& 
& $\surd$
& $\surd$
&
& 
& $e^2\frac{E^2}{\Lambda^2}\parallel e^2\frac{M_W^2}{E^2}$ 
& $\gamma \gamma \to W^- W^+ /(LL)$\\
LC($\gamma \gamma $)
& 
& $\triangle$
& $\triangle$
& $\triangle$
& 
& 
& $\triangle$
& $\triangle$
&
& 
& $e^2g\frac{Ef_\pi}{\Lambda^2}\parallel e^2\frac{M_W}{E}$  
& $\gamma\gamma \to W^-W^+ /(LT)$\\
& & & & & & & & & & & & \\
\hline\hline 
\end{tabular}}

\vspace{0.2cm}

\null\noindent
$\ell_4$ and $\ell_5$ at the 1.6 TeV LC has been carried out in Ref. 30.
The results are shown in Fig. 2 which shows that with polarized
electron beams, $\ell_4$ and $\ell_5$ can be determined at a higher accuracy.
Determination of custodial-symmetry-violating-term coefficients $\ell_6$ and 
$\ell_7$ via the interplay between the $V_LV_L$ fusion and $VVV$ production
has been studied in Ref. 31.

\centerline{\epsfig{figure=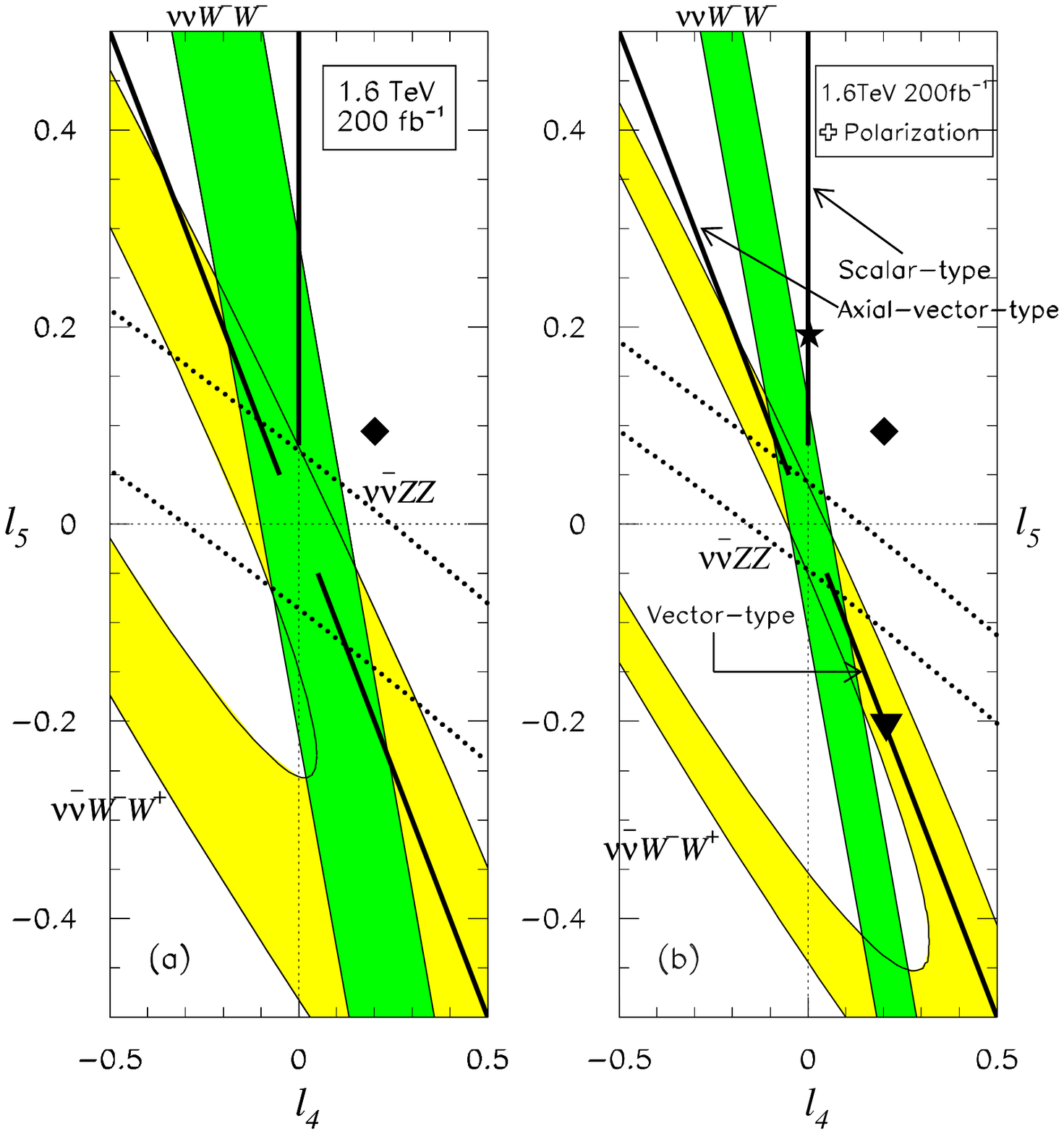,height=6cm}}
\vspace{-0.6cm}
\null\noindent{\footnotesize {\bf Fig. 2.} Determining the coefficients $\ell_4,~\ell_5$ at the 1.6 TeV
$e^+e^-/e^-e^-$ LC's. The $\pm 1\sigma$ exclusion contours are displayed.
(a) unpolarized case; (b) the case of $90\%(65\%)$ polarized $e^-(e^+)$ beam.
The thick solid lines are contributions from certain simple theoretical
models. Quoted from Ref. 30.}
\vspace{0.6cm}

Once the coefficients $\ell_n$s are measured at the LHC and the LC, the
next problem needed to solve is to study what underlying theory will
give rise to this set of coefficients. Only with this theoretical study the 
probe of the EWSB mechanism can be complete. Such a study is difficult
due to the nonperturbative nature, and there is no such kind of systematic 
study yet.

This kind of study is similar to the problem of deriving the Gasser-Leutwyler
Lagrangian for low lying pseudoscalar mesons (the {\it chiral
Lagrangian}) \cite{GL} from the fundamental theory of QCD. Very recently,
some progress in this case has been made in Ref. 33 in which
the Gasser-Leutwyler Lagrangian is formally derived from the first principles 
of QCD without taking approximations, and all the coefficients in the
Gasser-Leutwyler Lagrangian are expressed in terms of certain Green's
functions in QCD, which can be regarded as the QCD definitions of the
Gasser-Leutwyler coefficients. 
The method in Ref. 33 can be applied to the electroweak theory
to make the above desired study.

\vspace{1cm}
\begin{center}
{\bf VII. Conclusions}
\end{center}

Despite of the success of the SM, its EWSB sector is still not clear. The 
assumed elementary Higgs boson in the EWSM has not been found, and the
present LEP2 bound on the Higgs boson mass is $m_H>107.7$ GeV.
Since {\it the EWSB mechanism concerns the understanding of the origin of 
particle masses}, the probe of it is a very interesting and important topic 
in current particle physics. The SM Higgs sector suffers from the 
well-known problems of {\it triviality} and {\it unnaturalness}, so that the 
EWSB sector may concern new physics. From various analyses in Sec. 2,
we see that {\it the $Z$-pole precision data do not necessarily imply
the existence of a light Higgs boson}. So 
that the search for the Higgs boson should be carried out in the whole 
possible energy range up to 1 TeV. 
If a light Higgs boson (elementary or composite) exists, it can 
certainly be found, as we have seen, at the future high energy colliders such 
as the LHC, the LC (including the $\gamma\gamma$ and $e\gamma$ colliders), etc.
The LC has the advantage of low hadronic backgrounds. After finding the Higgs 
boson, we have to further study its properties to see if it is just the EWSM 
Higgs boson, or a Higgs boson in a more complicated new physics model (e.g. 
the MSSM), or it is composite.

If there is no light Higgs boson, the EWSB mechanism must
be strongly interacting. Some strongly interacting EWSB models contain
extra heavy gauge bosons below 1 TeV, and many strongly interacting
EWSB models contain certain pseudo Goldstone bosons (PGBs) in the few hundred 
GeV range characterizing the models. Therefore, a feasible way of probing the 
EWSN mechanism in this case is to test the extra gauge boson and PGB effects 
in certain processes at the high energy colliders, especially in top quark 
production processes. Another way of probing the EWSB mechanism, which is most 
direct but not easiest, is the study of the longitudinal weak boson reactions 
at high energy colliders. it is specially important if there is neither light 
Higgs boson nor a light resonance related to the EWSM mechanism below 1 TeV. 
We have seen that there can be a general model-independent probe of the EWSB
mechanism from measuring the coefficients in the EWCL via the study of 
longitudinal weak boson reactions. We have also seen that those coefficients 
can all be measured at the LHC and the LC, and for this purpose, the LHC alone 
is not enough.

In summary, particle physics will be in a crucial status of clarifying
the choice of different directions of new physics when we go to the TeV energy 
scale. The LHC and the LC will be important equipments for studying TeV 
physics and will help us to know to which direction we should further go. 

\null\noindent
{References}

\end{document}